\documentclass[12pt]{article}
\usepackage{bm}
\usepackage{graphicx}
\begin{document}
\title{PERTURBED INVARIANT UNDER A CYCLIC PERMUTATION WITH TRACE OF NEUTRINO MASS MATRIX REMAIN CONSTANT}
\date{}
\maketitle
\begin{center}
\textbf{Asan Damanik\footnote{E-mail: d.asan@lycos.com}}\\
\itshape Department of Physics, Faculty of Science and Technology,\\ Sanata Dharma University,\\ Kampus III USD Paingan, Maguwoharjo, Sleman,Yogyakarta, Indonesia\\
\end{center}
\abstract{We construct a neutrino mass matrix $M_{\nu}$ via a seesaw mechanism with perturbed invariant under a cyclic permutation by introducing one parameter $\delta$ into the diagonal elements of $M_{\nu}$ with assumption that trace of the perturbed $M_{\nu}$ is equal to trace of the unperturbed $M_{\nu}$.  We found that the perturbed neutrino mass matrices $M_{\nu}$ can predicts the mass-squared difference $\Delta m_{ij}^{2}\neq 0$ with the possible hierarchy of neutrino mass is normal or inverted hierarchy.  By using the advantages of the mass-squared differences and mixing parameters data from neutrino oscillation experiments, we then have neutrino masses in inverted hierarchy with masses: $\left|m_{1}\right|=0.101023$ eV, $\left|m_{2}\right|=0.101428$ eV, and $\left|m_{3}\right|=0.084413$ eV.}

\begin{flushleft}
PACs: 14.60.Pq
\end{flushleft}

\section{Introduction}

Recently, there is a convincing evidence that neutrinos have a tiny non-zero mass.  The evidence of neutrino mass is based on the experimental facts that both solar and atmospheric neutrinos undergo oscillations.\cite{Fukuda98}-\cite{Ahmad}  A global analysis of neutrino oscillations data gives the best fit value to solar neutrino mass-squared differences,\cite{Gonzales-Garcia04}
\begin{equation}
\Delta m_{21}^{2}=(8.2_{-0.3}^{+0.3})\times 10^{-5}~{\rm eV^2}~
 \label{11}
\end{equation}
with
\begin{equation}
\tan^{2}\theta_{21}=0.39_{-0.04}^{+0.05},
 \label{21}
\end{equation}
and for the atmospheric neutrino mass-squared differences
\begin{equation}
\Delta m_{32}^{2}=(2.2_{-0.4}^{+0.6})\times 10^{-3}~{\rm eV^2}~
 \label{22}
\end{equation}
with
\begin{equation}
\tan^{2}\theta_{32}=1.0_{-0.26}^{+0.35},
\end{equation}
where $\Delta m_{ij}^2=m_{i}^2-m_{j}^2~ (i,j=1,2,3)$ with $m_{i}$ is the neutrino mass in eigenstates basis $\nu_{i}~(i=1,2,3)$, and $\theta_{ij}$ is the mixing angle between $\nu_{i}$ and $\nu_{j}$.  The mass eigenstates basis are related to the weak (flavor) eigenstates basis $(\nu_{e},\nu_{\mu},\nu_{\tau})$ as follows,
\begin{equation}
\bordermatrix{& \cr
&\nu_{e}\cr
&\nu_{\mu}\cr
&\nu_{\tau}\cr}=V\bordermatrix{& \cr
&\nu_{1}\cr
&\nu_{2}\cr
&\nu_{3}\cr}
 \label{5}
\end{equation}
where $V$ is the mixing matrix.

It is also known that neutrino masses are very small compared to its corresponding charged lepton masses and mixing does exist in neutrino sector.  Charged lepton mass has a normal hierarchy, but neutrino mass can have either a normal or an inverted hierarchy.  Thus, neutrinos have some different properties from charged leptons.  From the theoretical side, it has been a guiding principle that the presence of hierarchies or of tiny quantities imply a certain protection symmetry in underlying physics.  The candidates of such symmetry in neutrino physics may include $U(1)_{L'}$ based on the conservation of $L_{e}-L_{\mu}-L_{\tau}=L'$ and a $\mu-\tau$ symmetry based on the invariance of flavor neutrino mass term underlying the interchange of $\nu_{\mu}$ and $\nu_{\tau}$.

To accommodate a tiny non-zero neutrino mass that can produce the mass-squared differences and the neutrino mixing, several models for the neutrino mass matrices together with the responsible mechanisms for generating it patterns have been proposed by many authors.  One of the interesting mechanism which can generate a small neutrino mass is the seesaw mechanism, in which the right-handed neutrino $\nu_{R}$ has a large Majorana mass $M_{N}$ and the left-handed neutrino $\nu_{L}$ obtain a mass through leakage of the order of $~(m/M_{N})$ with $m$ is the Dirac mass.\cite{Fukugita03}

According to seesaw mechanism,\cite{Gell-Mann79} the neutrino mass matrix $M_{\nu}$ is given by,
\begin{eqnarray}
M_{\nu}\approx -M_{D}M_{N}^{-1}M_{D}^T
 \label{Mnu}
\end{eqnarray}
where $M_{D}$ and $M_{N}$ are the Dirac and Majorana mass matrices respectively.  The mass matrix model of a massive Majorana neutrino $M_{N}$ which is constrained by the solar and atmospheric neutrinos deficit and incorporating the seesaw mechanism and Peccei-Quinn symmetry have been reported by Fukuyama and Nishiura.\cite{Fukuyama97}  Neutrino mass matrix patterns together with its underlying symmetry become an interesting research topic during the last few years.  In related to the seesaw mechanism, Ma\cite{Ma05} pointed out that it is more sense to consider the structure of $M_{N}$ for its imprint on $M_{\nu}$.

In order to consider the structure of the $M_{N}$ for its imprint on $M_{\nu}$, in this paper we construct the neutrino mass matrices $M_{\nu}$ arise from a seesaw mechanism with both heavy Majorana and Dirac neutrino mass matrix are invariant under a cyclic permutation.  As we have already knew that neutrino mass matrix which is invariant under a cyclic permutation gives $m_{1}=m_{3}$ and then it fails to predict mass-squared difference $\Delta m_{31}^{2}\neq 0$.  The charged-lepton mass matrix which is invariant under a cyclic permutation have been analyzed by Koide\cite{Koide05} that also suggested to break the invariant under a cyclic permutation if we want to obtain the charged-lepton mass spectrum compatible with the empirical fact.

To overcome the weakness of cyclic permutation on predicting mass-squared differences, in this paper we introduce a perturbation into neutrino mass matrix with the assumption that the perturbed cyclic permutation mass matrix has the same trace with the unperturbed neutrino mass matrix.  This paper is organized as follows: In Section 2, we construct the heavy Majorana and Dirac neutrino mass matrices which are invariant under a cyclic permutation.  The resulted neutrino mass matrices to be used for obtaining the neutrino mass matrix $M_{\nu}$ in the scheme of seesaw mechanism.  In Section 3, we use a seesaw mechanism for obtaining neutrino mass matrix and evaluate its phenomenological consequences.  Finally, the Section 4 is devoted to a conclusion.

\section{Neutrino Mass Matrix with Invariant under Cyclic Permutation}

As we have already stated above, the aim of this paper is to study the phenomenological consequences of the perturbed cyclic permutation neutrino mass matrices arise from a seesaw mechanism with both heavy Majorana and Dirac neutrino mass matrices are invariant under a cyclic permutation.  The seesaw mechanism to be considered in this paper is the type-I seesaw.  In order to realize the goals of this section, first we write down the the possible patterns for heavy Majorana neutrino mass matrices $M_{N}$ is invariant under a cyclic permutation.  Second, we write down the possible patterns for Dirac neutrino mass matrices by taking into account the same constraints that we have imposed on heavy Majorana neutrino mass matrix.
  
We consider the Majorana neutrino mass matrix $M_{N}$ in Eq.~(\ref{Mnu}) to be symmetric in form and that matrix is given by
\begin{equation}
M_{N}=\bordermatrix{& & &\cr
&A &B &C\cr
&B &D &E\cr
&C &E &F\cr}.
 \label{MN}
\end{equation}

In order to obtain the $M_{N}$ that invariant under a cyclic permutation among neutrino fields: $\nu_{1}\rightarrow \nu_{2}\rightarrow \nu_{3}\rightarrow \nu_{1}$, first we define
\begin{equation}
\nu'_{i}=U_{ij}\nu_{j},
 \label{T}
\end{equation}
where $U_{ij}$ are the entries of the cyclic permutation matrix $U$.  From Eq.~(\ref{T}), one can see that the $M_{N}$ matrix satisfy
\begin{equation}
M'_{N}=U^{T}M_{N}U.
 \label{M}
\end{equation}
If the $M_{N}$ matrix is invariant under a cyclic permutation, then the pattern of the $M'_{N}$ is the same with the $M_{N}$ pattern.

By imposing the requirement that the form of the $M_{N}$ matrix in Eq. (\ref{MN}) must be invariant under a cyclic permutation together with the requirement that the $M_{N}$ is non-singular matrix such that the $M_{N}$ has a $M_{N}^{-1}$, then we have three possible patterns for heavy Majorana neutrino mass matrix $M_{N}$ as follow
\begin{eqnarray}
M_{N}=\bordermatrix{& & &\cr
&A &B &B\cr
&B &A &B\cr
&B &B &A\cr},
 \label{MN11}\\
M_{N}=\bordermatrix{& & &\cr
&A &0 &0\cr
&0 &A &0\cr
&0 &0 &A\cr},
 \label{MN12}\\
M_{N}=\bordermatrix{& & &\cr
&0 &B &B\cr
&B &0 &B\cr
&B &B &0\cr}.
 \label{MN13}
\end{eqnarray} 
From Eqs.~(\ref{MN11}),~(\ref{MN12}), and~(\ref{MN13}), one can see that the patterns of neutrino mass matrices in Eqs.~(\ref{MN12}) and (\ref{MN13}) are special cases of the neutrino mass matrix pattern in Eq.~(\ref{MN11}).  Thus, the neutrino mass matrix given by Eq.~(\ref{MN11}) is the most general pattern, and we will consider it as a good candidate for neutrino mass matrix $M_{N}$.  

To obtain the neutrino mass matrices $M_{\nu}$ arise from a seesaw mechanism (using Eq.~(\ref{Mnu})), we should know the patterns of the Dirac neutrino mass matrices $M_{D}$.  Because the heavy neutrino fields are part of the Dirac mass term, according to Eqs.~(\ref{T}) and (\ref{M}), the possible patterns for Dirac neutrino mass matrices are given by
\begin{eqnarray}
M_{D}=\bordermatrix{& & &\cr
&a &a &a\cr
&a &a &a\cr
&a &a &a\cr},
 \label{MD0}\\
M_{D}=\bordermatrix{& & &\cr
&a &b &b\cr
&b &a &b\cr
&b &b &a\cr},
 \label{MD1}\\
M_{D}=\bordermatrix{& & &\cr
&a &0 &0\cr
&0 &a &0\cr
&0 &0 &a\cr},
 \label{MD2}\\
M_{D}=\bordermatrix{& & &\cr
&0 &b &b\cr
&b &0 &b\cr
&b &b &0\cr}.
 \label{MD3}
\end{eqnarray}
It is apparent from Eqs.~(\ref{MD0})-(\ref{MD3}) that the pattern of neutrino mass matrix  $M_{D}$ in Eq. (\ref{MD1}) is the most general pattern.  Thus, we will consider neutrino mass matrix $M_{D}$ in Eq.~(\ref{MD1}) as a good candidate for Dirac neutrino matrix.

\section{Neutrino Mass matrix via a Seesaw Mechanism}

Using the seesaw mechanism in Eq.~(\ref{Mnu}), the heavy Majorana neutrino mass matrix in Eq.~(\ref{MN11}), and the Dirac neutrino mass matrix in Eq.~(\ref{MD1}), we then obtain a neutrino mass matrix with pattern,
\begin{eqnarray}
M_{\nu}=\bordermatrix{& & &\cr
&P &Q &Q\cr
&Q &P &Q\cr
&Q &Q &P\cr}.
 \label{MD01}
\end{eqnarray} 
The eigenvalues of the neutrino mass matrix in Eq.~(\ref{MD01}) are given by
\begin{eqnarray}
\lambda_{1}=\lambda_{2}=P-Q,\ \lambda_{3}=P+2Q.
 \label{as}
\end{eqnarray}

It is easy to see that one of the eigenvectors of the $M_{\nu}$ is $(1,1,1)^{T}$ and this eigenvector corresponds to eigenvalue $\lambda_{3}$.  Thus, the eigenvalue $\lambda_{3}$ should be identified as neutrino mass $m_{2}$, meanwhile $\lambda_{1}$ and $\lambda_{2}$ correspond to neutrino masses $m_{1}$ and $m_{2}$.  Finally, the neutrino mass matrix in Eq.~(\ref{MD01}) gives neutrino masses,
\begin{equation}
m_{1}=m_{3}=P-Q,\ m_{2}=P+2Q.
 \label{as1}
\end{equation}
From Eq.~(\ref{as1}), we have $\Delta m_{21}^{2}= \left|\Delta m_{32}^{2}\right|=6PQ+3Q^{2}$ which is contrary to the experimental fact.  Thus, the neutrino mass matrix in Eq.~(\ref{MD01}) could not reproduce the mass-squared difference $\Delta m_{21}^{2}<< \Delta m_{32}^{2}$.  It is also apparent that the resulted neutrino mass matrix in this scenario gives $m_{1}+m_{2}+m_{3}=3P$ is which is equal to $Tr(M_{\nu})$.

Even though neutrino mass matrix that invariant under a cyclic permutation, as one can see in Eq.~(\ref{MD01}), could not predict correctly the experimental data, we can still use it as a neutrino mass matrix.  In order to obtain neutrino mass matrix that can give correct predictions on mass-squared differences and mixing parameters, we modify the neutrino mass matrix $M_{\nu}$ in Eq.~(\ref{MD01}) by introducing one parameter $\delta$ to perturb the diagonal elements of $M_{\nu}$ such that the perturbed mass matrix satisfies the requirement $Tr(M_{\nu})=3P$.  In this scenario, we then can put the neutrino mass matrix $M_{\nu}$ in form,
\begin{equation}
M_{\nu}=\bordermatrix{& & &\cr
&P+2\delta &Q &Q\cr
&Q &P-\delta &Q\cr
&Q &Q &P-\delta\cr}.
 \label{MD011}
\end{equation} 

The eigenvalues of the neutrino mass matrix in Eq.~(\ref{MD011}) read
\begin{eqnarray}
\beta_{1,2}=P+\frac{Q}{2}+\frac{\delta}{2}\mp\frac{\sqrt{9\delta^{2}-6Q\delta+9Q^{2}}}{2},\\
\beta_{3}=P-Q-\delta.
 \label{ec}
\end{eqnarray}
If the neutrino mass matrices $M_{\nu}$ in Eq.~(\ref{MD011}) is diagonalized by mixing matrix $V$ in Eq.~(\ref{5}) with $V$ given by\cite{Ma05}
\begin{equation}
V=\bordermatrix{& & &\cr
&\cos\theta &-\sin\theta &0\cr
&\sin\theta/\sqrt{2} &\cos\theta/\sqrt{2} &-1/\sqrt{2}\cr
&\sin\theta/\sqrt{2} &\cos\theta/\sqrt{2} &1/\sqrt{2}\cr},
 \label{511}
\end{equation}
then we obtain,
\begin{equation}
\tan^{2}(2\theta)=\frac{8Q^{2}}{(Q-3\delta)^{2}},
 \label{Teta}
\end{equation}
and neutrino masses as follow,
\begin{eqnarray}
m_{1}=P+\frac{Q}{2}+\frac{\delta}{2}-\frac{\sqrt{9\delta^{2}-6Q\delta+9Q^{2}}}{2},
 \label{m1}\\
m_{2}=P+\frac{Q}{2}+\frac{\delta}{2}+\frac{\sqrt{9\delta^{2}-6Q\delta+9Q^{2}}}{2},
 \label{m2}\\
m_{3}=P-Q-\delta.
 \label{m3}
\end{eqnarray}
One can see that the obtained neutrino masses in this scenario is an inverted hierarchy with masses: $\left|m_{3}\right|<\left|m_{1}\right|<\left|m_{2}\right|$.

If $\theta$ is the $\theta_{21}$ in Eq.~(\ref{21}), then from Eq.~(\ref{Teta}) we have $\delta=-0.1271Q$. If we insert this $\delta$ value into Eqs.~(\ref{m1})-(\ref{m3}), then we have the neutrino masses as follow
\begin{eqnarray}
m_{1}=P-0.1374Q,
 \label{m11}\\
m_{2}=P+2.0103Q,
 \label{m22}\\
m_{3}=P-0.8729Q.
 \label{m33}
\end{eqnarray}
The plot of $m_{1}$, $m_{2}$, and $m_{3}$ as function of parameters $P$ and $Q$ are shown in Fig.1.

\begin{figure}[h]
 \centering
 \includegraphics[width=0.3\textwidth]{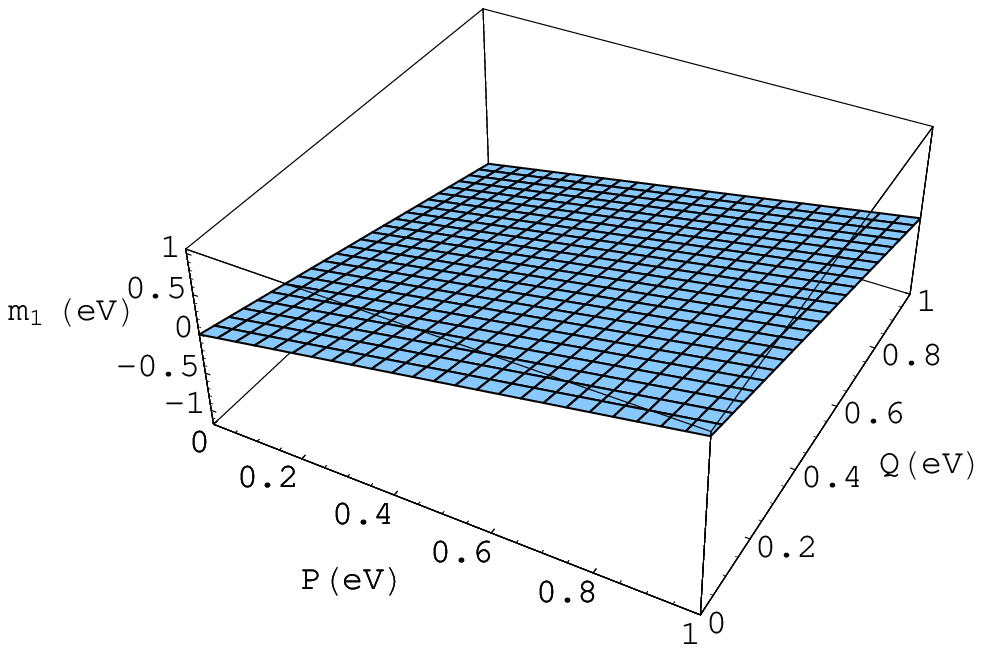}
 \includegraphics[width=0.3\textwidth]{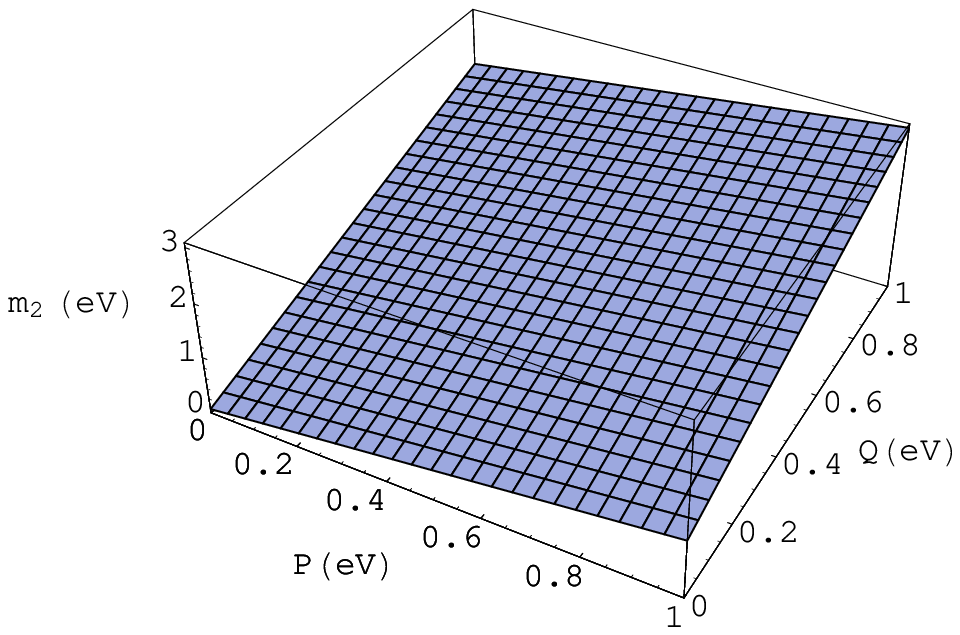}
 \includegraphics[width=0.3\textwidth]{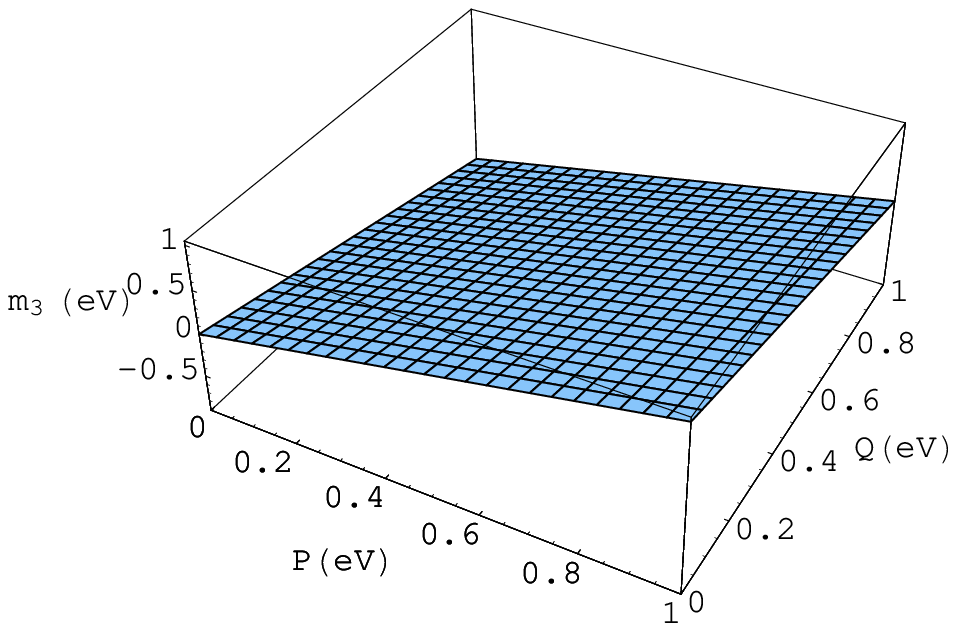}
  \caption{Neutrino masses $m_{1}$, $m_{2}$, and $m_{3}$ as function of parameters $P$ and $Q$.}
	\label{figure1}
\end{figure}

From Fig.~1 we can see that the neutrino mass can have normal, degenerate, or inverted hierarchy which it depends on the sign and values of of parameters $P$ and $Q$.  For example, if we put the values of $P=-0.8$ and $Q=1$, then we have neutrino masses in normal hierarchy: $\left|m_{1}\right|<\left|m_{2}\right|<\left|m_{3}\right|$.  The degenerate hierarchy: $\left|m_{1}\right|\approx\left|m_{2}\right|\approx\left|m_{3}\right|$  can be obtained if we put the value of parameter $Q\approx 0$, and the inverted hierarchy: $\left|m_{3}\right|<\left|m_{1}\right|<\left|m_{2}\right|$ is produced for $P>0.8729Q$ and $Q>0$.

If we use the advantages of the experimental data of neutrino oscillation in Eqs.~(\ref{11}) and (\ref{22}), from Eqs.~(\ref{m1}),~(\ref{m2}), and~(\ref{m3}), then we obtain the neutrino masses,
\begin{equation}
\left|m_{1}\right|=0.101023 ~{\rm eV}~, \; \left|m_{2}\right|=0.101428 ~{\rm eV}~,\;\left|m_{3}\right|=0.084413 ~{\rm eV}, 
 \label{NM}
\end{equation}
for $Q=-0.06432$ eV and $P=0.02827$ eV.  One can see that the value of $\delta=0.008176$ eV is smaller than the values of $P$ and $Q$.  Inserting the obtained values of $P, Q$, and $\delta$ into Eq.~(\ref{MD011}), we finally have neutrino mass matrix in eV unit as follow
\begin{equation}
M_{\nu}=\bordermatrix{& & &\cr
&0.04463 &-0.06432 &-0.06432\cr
&-0.06432 &0.02010 &-0.06432\cr
&-0.06432 &-0.06432 &0.02010\cr}.
\end{equation}

\section{Conclusion}
Neutrino mass matrix $M_{\nu}$ arise from a seesaw mechanism, with both heavy Majorana and Dirac neutrino mass matrices are invariant under a cyclic permutation, can not be used to explain the present experimental data of neutrino oscillation.  By introducing one parameter $\delta$ to perturb the diagonal elements of $M_{\nu}$ with the assumption that the value of the trace of $M_{\nu}$ remain constant, we then have a neutrino mass matrix that can be used to explain mass-squared differences.  In this scenario, the possible hierarchy of neutrino mass that can be used to explain mass-squared differences is normal or inverted hierarchy.  By using the mass-squared differences and mixing parameters which obtained from neutrino oscillation experiments, we then have neutrino masses in inverted hierarchy with masses: $\left|m_{1}\right|=0.101023$ eV, $\left|m_{2}\right|=0.101428$ eV, and $\left|m_{3}\right|=0.084413$ eV.

\section*{Acknowledgments}
Author would like to thank reviewers for their useful comments and suggestions.


\begin{thebibliography}{00}
\bibitem{Fukuda98}
Y. Fukuda {\it et al.}, {\it Phys. Rev. Lett.} {\bf81}, 1158 (1998).
\bibitem{Fukuda99}
 Y. Fukuda {\it et al.}, {\it Phys. Rev. Lett}. {\bf82}, 2430 (1999).
\bibitem{Ahn}
M. H. Ahn {\it et al.}, {\it Phys. Rev. Lett}. {\bf90}, 041801-1 (2003).
\bibitem{Toshito}
T. Toshito {\it et al.}, hep-ex/0105023.
\bibitem{Giacomelli}
G. Giacomelli and M. Giorgini, hep-ex/0110021.
\bibitem{Ahmad}
Q.R. Ahmad {\it et al.}, {\it Phys. Rev. Lett.} {\bf89}, 011301 (2002).
\bibitem{Gonzales-Garcia04} 
M.C. Gonzales-Garcia, {\it Phys. Scripta} {\bf T121}, 72 (2005).
\bibitem{Fukugita03}
M. Fukugita and T. Yanagida, {\it Physics of Neutrinos and Application to Astrophysics} (Springer-Verlag, 2003).
\bibitem{Gell-Mann79}
M. Gell-Mann, P. Ramond, and R. Slansky, {\it Supergravity}, eds. P. van Niewenhuzen and D. Z. Freedman  (North-Holland, 1979).
\bibitem{Fukuyama97}
T. Fukuyama and H. Nishiura, arXiv: hep-ph/9702253.
\bibitem{Ma05}
E. Ma, {\it Phys. Rev.} {\bf D71}, 111301 (2005).
\bibitem{Koide05}
Y. Koide, arXiv: hep-ph/0005137v1.
\end{thebibliography}
\end{document}